\documentclass[onecolumn,prd,nofootinbib,showpacs,floatfix]{revtex4}
\usepackage{epsfig}
\usepackage{color}
\usepackage{amsmath}
\newcommand{\be}{\begin{equation}}
\newcommand{\ee}{\end{equation}}
\newcommand{\bea}{\begin{eqnarray}}
\newcommand{\eea}{\end{eqnarray}}
\newcommand{\eqn}[1]{(\ref{#1})}
\newcommand{\pth}{\texttt{PArthENoPE} }
\newcommand{\pthnew}{\texttt{PArthENoPE 2.0} }
\newcommand{\dneff}{\Delta N_{\rm eff}}
\newcommand{\pp}{~~~.}
\newcommand{\vv}{~~~,}
\newcommand{\p}{{\rm p}}
\newcommand{\lrt}{\leftrightarrow}
\newcommand{\rt}{\rightarrow}
\newcommand{\pe}{{\phi_e}}
\newcommand{\hrho}{\hat{\rho}}
\newcommand{\rhol}{\rho_\Lambda}
\newcommand{\lh}{\hat{L}}
\newcommand{\ds}{\displaystyle}
\newcommand{\hH}{\widehat{H}}
\newcommand{\hp}{\hat{\rm p}}

\newcommand{\hdmi}{\Delta \widehat{M}_i}

\newcommand{\hgi}{\widehat{\Gamma}_i}

\newcommand{\hnb}{\hat{n}_B}
\newcommand{\tgi}{\widetilde{\Gamma}_i}

\begin{document}

\title{\pth reloaded}

\author{R. Consiglio$^{1}$, P.\ F.\ de Salas$^{2}$, G. Mangano$^{3}$, G. Miele$^{1,3}$, S.\ Pastor$^{2}$, and O. Pisanti$^{1,3}$\footnote{Corresponding author. Ofelia Pisanti, e-mail: pisanti@na.infn.it}}

\affiliation{ $^1$Dipartimento di Fisica E. Pancini, Universit\`a di Napoli Federico II, Via Cintia, I-80126 Napoli, Italy,}
\affiliation{$^2$Instituto de F\'{\i}sica Corpuscular  (CSIC-Universitat de Val\`{e}ncia),
c/ Catedr\'atico Jos\'e Beltr\'an, 2, 46980 Paterna (Valencia), Spain,}
\affiliation{ $^3$INFN, Sezione di Napoli, Via Cintia, I-80126 Napoli, Italy.}

\begin{abstract}
{We describe the main features of a new and updated version of the program {\texttt{PArthENoPE}}, which computes the abundances of light elements produced during Big Bang Nucleosynthesis. As the previous first release in 2008, the new one, {\texttt{PArthENoPE 2.0}}, will be soon publicly available and distributed from the code site, {\texttt{http://parthenope.na.infn.it}}.  Apart from minor changes, which will be also detailed, the main improvements are as follows. The powerful, but not freely accessible, NAG routines have been substituted by ODEPACK libraries, without any significant loss in precision. Moreover, we have developed a Graphical User Interface (GUI) which allows a friendly use of the code and a simpler implementation of running for grids of input parameters. Finally, we report the results of \pthnew for a minimal BBN scenario with free radiation energy density.}
\\ \\
{\bf NEW VERSION PROGRAM SUMMARY}

\begin{small}
\noindent
\\
{\em Program Title:}  \pthnew \\
{\em Licensing provisions:} GPLv3 \\
{\em Programming language:} Fortran 77 and Python \\
{\em Supplementary material:} User Manual available on the web page {\texttt{http://parthenope.na.infn.it}} \\
{\em Journal reference of previous version:} Comput.\ Phys.\ Commun.\ {\bf 178} (2008) 956-971 \\
{\em Does the new version supersede the previous version?:} Yes \\
{\em Reasons for the new version:} Make the code more versatile and user friendly \\
{\em Summary of revisions:} 1) Publicly available libraries 2) GUI for configuration \\
{\em Nature of problem(approx. 50-250 words):}  Computation of yields of light elements synthesized in the primordial universe\\
{\em Solution method(approx. 50-250 words):} Livermore Solver for Ordinary Differential Equations (LSODE) for stiff and nonstiff systems\\
{\em Computers}: PC-compatible running Fortran on Unix/Linux or Mac; Python GUI for Unix/Linux or Mac\\
{\em Program obtainable from}: web page {\texttt{http://parthenope.na.infn.it}}\\
{\em E-mail}: parthenope@na.infn.it\\
\\
\end{small}

\end{abstract}


\maketitle

\section{Introduction}
\label{s:intr}

Although during the last decade the major improvements in our understanding of cosmology came from high-precision measurements of Cosmic Microwave Background (CMB) anisotropies \cite{Hinshaw:2012aka,Ade:2015xua}, Big Bang Nucleosynthesis (BBN) remains a fundamental pillar and a useful tool to test the overall compatibility of the cosmological model. BBN also provides a way to constrain more exotic scenarios involving new physics beyond the Standard Model of fundamental interactions when the universe was only few seconds up to few minutes old. 

As well known, when the temperature of the primordial plasma reached  a value of $\sim$ 1 MeV,  light nuclides like $^2$H, $^3$He, $^4$He and, to a smaller extent, $^7$Be and $^7$Li, started to be produced through a complicated network of nuclear processes. During this period, the early universe behaved as a nuclear reactor, leaving as a final product a spectrum of values for the relative abundances of these nuclear ``ashes'' with respect to hydrogen, spanning a large range of different orders of magnitude. Such theoretical primordial values can be compared with the astrophysical observations performed in peculiar astrophysical environments that are characterized by a small stellar contamination. For a  standard cosmological scenario and in the framework of the electroweak Standard Model, BBN is characterized by a single free parameter,  the baryon to
photon number density, which can be deduced by comparing theoretical predictions with the observations. Remarkably, the value determined in this way using deuterium and $^4$He abundances is in good agreement with  that obtained from the analysis of CMB data \cite{Ade:2015xua}. There is still a long standing $^7$Li problem, whose theoretical prediction is a factor 2-3 higher than astrophysical measurements, but this disagreement can be due to stellar contamination affecting the value measured in the Spite plateau as suggested by a strong positive correlation with Fe/H \cite{Sbordone:2010zi}.

Since the concordance for the standard cosmological scenario is quite satisfactory, this means that possible effects on BBN due to new physics should be quite small, and comparing theoretical expectations with astrophysical data allows for a quantitative study of extensions of the $\Lambda$CDM model. One particularly simple example of an extended model includes new light degrees of freedom which might contribute to the total cosmological energy density in the form of radiation, in addition to photons and the three active neutrinos, typically encoded in the parameter generally known as the {\it effective number of neutrinos}, $N_{\rm eff}$, see e.g. \cite{Iocco:2008va}. Maybe, at the time of writing, the more physically motivated candidate which might be produced in the early universe is a light sterile neutrino with an order eV mass, as suggested in order to explain several anomalies in the data from neutrino oscillation experiments \cite{Abazajian:2012ys}.

A faithful use of primordial nucleosynthesis as a {\it model tester} has required an intense work to improve the level of accuracy of BBN predictions, at least up to the level of experimental uncertainties. To pursue such a goal, in the last two decades a careful analysis of several key aspects of the physics involved in BBN has been performed. This started from an improvement of the accuracy of the expression for weak reactions keeping neutrons and protons in chemical equilibrium \cite{Lopez:1998vk,Esposito:1998rc,Esposito:1999sz,Esposito:2000hh,Serpico:2004gx}. The same was done for the neutrino decoupling process, carefully studied by several authors by explicitly solving the corresponding kinetic equations \cite{Hannestad:1995rs,Dolgov:1997mb,Mangano:2001iu,Mangano:2005cc,Birrell:2014uka,deSalas:2016ztq,Grohs:2017iit}. These two improvements allowed to reduce the theoretical uncertainty on the $^4$He mass fraction down to  0.1\%, due to the experimental uncertainty on the neutron lifetime, $\tau_n= (880.2 \pm 1.0)$ s \cite{PDG}. Finally, an exhaustive study of the whole nuclear chain entering the BBN network was performed in \cite{Serpico:2004gx,Cyburt:2004cq} taking advantage of the NACRE Collaboration database \cite{nacre} and NACRE II \cite{Xu:2013fha}. This analysis consisted, for each reaction, in an update of its expression and uncertainty taking into account possible new experimental measurements, and, finally, in the calculation of the corresponding thermal averaged rates.  

All these steps were implemented in a new public BBN code, $\pth$ \cite{parthenope},  which updated the pioneering achievements of 
\cite{Wagoner:1966pv,Wagoner2,Wagoner:1972jh,Kawano:1988vh,Smith:1992yy}. After almost ten years from the first release of the $\pth$ code it was mandatory to improve its performance with a new version, \texttt{PArthENoPE 2.0}, overcoming several critical features of the previous one. In particular, we have developed a Graphical User Interface (GUI) allowing a more friendly use of the code and replaced the NAG libraries, particularly suitable to solve complex stiff numerical problems, with more portable libraries without losing on the precision level.

In the present paper we  give a brief summary of the physics of BBN in Section \ref{s:equations} describing the theoretical framework and reporting the set of equations solved by the code and the free parameters involved. In Section \ref{new release} we present the characteristics of the new release, by enlightening the advantages of this new version of the code. In Section \ref{s:compare} we compare the performances of $\pthnew$ with respect to \texttt{PArthENoPE}, whereas in Section \ref{s:analysis} we present, as an example of the use
of the new code, a BBN analysis in order to determine the allowed ranges of the baryon to photon number density in the standard case and in an extended model with free radiation energy density. Finally, we present our conclusions in Section \ref{s:concl}.

%%%%%%%%%%%%%%%%%%%%%%%%%%%%%%%%%%%%%%%%%%%%%%%%%%

\section{The physics of BBN}
\label{s:equations}

We address the reader to Sec.\ 2.1 of \cite{parthenope} for a detailed description of the BBN set of equations and we give here only a brief summary. In the following we use natural units,  $\hbar=c=k_B=1$. 

We consider $N_{\rm nuc}$ species of nuclides, whose number densities $n_i$ are normalized with respect to the total number density of
baryons $n_B$,
\be
X_i=\frac{n_i}{n_B} \quad\quad\quad i=n,\,p,\,^2{\rm H}, \, ...
\pp
\ee
The set of differential equations ruling primordial nucleosynthesis is the following (see for example \cite{Wagoner:1966pv,Wagoner2,Wagoner:1972jh,Esposito:1999sz,Esposito:2000hh}):
\bea
&&\frac{\dot{a}}{a}  = H = \sqrt{\frac{8\, \pi G_{\rm N}}{3}~ \rho} \vv
\label{e:drdt} \\
&&\dot{n}_B = -\, 3\, H\, n_B \vv
\label{e:dnbdt} \\
&&\dot{\rho} = -\, 3 \, H~ (\rho + \p) \vv
\label{e:drhodt} \\
&&\dot{X}_i = \sum_{j,k,l}\, N_i \left(  \Gamma_{kl \rt ij}\,
\frac{X_l^{N_l}\, X_k^{N_k}}{N_l!\, N_k !}  \; - \; \Gamma_{ij \rt
kl}\, \frac{X_i^{N_i}\, X_j^{N_j}}{N_i !\, N_j  !} \right) \equiv
\Gamma_i \vv
\label{e:dXdt} \\
&& n_B~ \sum_j Z_j\, X_j =n_{e^-}-n_{e^+}\equiv
L \left(\frac{m_e}{T}, \pe\right)  \equiv
T^3~ \lh \left(\frac{m_e}{T}, \pe\right) \vv \label{e:charneut}
\eea
where $a(t)$ is the cosmological scale factor, $G_{\rm N}$ is the Newton constant and $\rho$ and $\p$ denote the total energy density and pressure,
respectively,
\bea
\rho &=& \rho_\gamma + \rho_e + \rho_\nu + \rho_B \vv \\
\p &=& \p_\gamma + \p_e + \p_\nu + \p_B\pp
\eea
In the previous equations $H$ is the Hubble parameter, $T$ the photon temperature, $i,j,k,l$ denote nuclear species, $N_i$ the number of nuclides of type $i$ entering a given reaction (and analogously $N_j$, $N_k$, $N_l$), $Z_i$ is the charge number of the $i-$th nuclide. The $\Gamma$'s denote symbolically the reaction rates, $\pe$ is the electron degeneracy parameter defined in terms of the electron chemical potential by $\pe\equiv\mu_e/T$. Finally, the function $\lh(\xi,\omega)$ is defined
as
\be
\lh(\xi,\omega) \equiv \frac{1}{\pi^2} \int_\xi^\infty
d\zeta~\zeta\, \sqrt{\zeta^2-\xi^2}~ \left(
\frac{1}{e^{\zeta-\omega}+1} - \frac{1}{e^{\zeta+\omega}+1}
\right) \pp \label{e:lfunc}
\ee

We do not consider neutrino oscillations, whose effect has been studied and shown to be sub-leading in \cite{Mangano:2005cc}. The reader can find further details on the neutrino decoupling stage in \cite{Serpico:2004gx,Mangano:2005cc,deSalas:2016ztq}. 

The set of equations~\eqn{e:drdt}-\eqn{e:charneut} are transformed into the set of differential equations implemented in \pthnew (see Appendix \ref{ap:erratum} for details and the definition of the quantities involved):
\be
\frac{d\pe}{dz} = \frac1z \frac{\lh\, \kappa_1 +\left(\hrho_{e
\gamma} + \hp_{e \gamma B}+
\frac{\mathcal{N}(z)}{3}\right)\,\kappa_2}{\lh\, \frac{\partial
\hrho_e}{ \partial \pe} -\frac{\partial \lh}{
\partial \pe} \left(\hrho_{e \gamma} + \hp_{e \gamma B}+
\frac{\mathcal{N}(z)}{3}\right)}\vv
\label{e:basic1a}
\ee
\be
\frac{dX_i}{d z}= -\frac{\hgi}{3\,
z\,\hH}\, \frac{\kappa_1 \, \frac{\partial \lh}{\partial \pe}
+\kappa_2 \, \frac{\partial \hrho_e}{\partial
\pe}}{\lh\,\frac{\partial\hrho_e}{\partial\pe}-\frac{\partial
\lh}{\partial \pe}\left(\hrho_{e \gamma} + \hp_{e \gamma B} +
\frac{\mathcal{N}(z)}{3}\right) }\vv
\label{e:basic2a}
\ee
which allow to follow the evolution of the $N_{\rm nuc}+1$ unknown quantities $(\pe,~ X_i)$ as functions of the dimensionless variable $z\equiv m_e/T$.

Equations~\eqn{e:basic1a} and~\eqn{e:basic2a} are solved by imposing the following initial conditions at $z_{\rm in}= m_e/(10\, {\rm MeV})$: 
\bea
\pe (z_{\rm in}) &=& \pe^0 \vv \\
X_1 (z_{\rm in}) & \equiv & X_n (z_{\rm in}) = \left(\exp\{\hat{q}\,
z_{\rm in}\}+1\right)^{-1} \vv \\
X_2 (z_{\rm in}) & \equiv & X_p(z_{\rm in}) = \left(\exp\{-\hat{q}\,
z_{\rm in}\}+1\right)^{-1} \vv \\
X_3 (z_{\rm in}) & \equiv & X_{^2{\rm H}}(z_{\rm in}) = g_{^2{\rm H}}~ 
\frac{4\,\zeta(3)}{\sqrt{\pi}}\, \left(
\frac{m_e}{M_N z_{\rm in}} \right)^{\frac{3}{2}}
\eta_i\, X_p(z_{\rm in})\, X_n(z_{\rm in}) \, \exp \left\{ \hat{B}_{^2{\rm H}} \,
z_{\rm in} \right\} \\
X_i (z_{\rm in}) &=& X_{\rm min} \quad\quad\quad\quad\quad i= \,^3{\rm H}, ... \pp
\eea
In the previous equations $\pe^0$ is the solution of the implicit equation
\be
\lh (z_{\rm in},\, \pe^0) =
\frac{2\, \zeta(3)}{\pi^2}~ \eta_i~ \sum_i Z_i\, X_i (z_{\rm in}) \vv
\label{e:L}
\ee
where $\eta_i\simeq 2.73\eta_f$ is the initial value of the baryon to photon number density ratio at $T=10\,{\rm MeV}$ (for a discussion of how it is related to the final value $\eta_f$ after the $e^+e^-$ annihilation stage see e.g.\ Section 4.2.2 in \cite{Serpico:2004gx}). Moreover, $\hat{q}=(M_n-M_p)/m_e$ and $\hat{B}_{^2{\rm H}}$ is the binding energy of deuterium normalized to the electron mass. Also note that nuclides with mass number larger than deuterium have their initial abundances equal to the numerical zero assumed in \pthnew ($X_{\rm min}$, corresponding to the variable YMIN, whose default setting is $10^{-30}$).

The nuclear processes considered in \pth and \pthnew are listed in detail in \cite{parthenope} (see \cite{Serpico:2004gx} for the relevant derivation of thermally averaged nuclear rates from experimental reaction data). We remind that the user can choose among three networks (small, intermediate and complete), which correspond to follow only the lighter nuclides, $^2$H, $^3$H, $^3$He, $^4$He, $^6$Li, $^7$Li, and $^7$Be, or enlarge to heavier nuclides. We stress here that already in the original version released in 2008 the rates of the reactions $^2$H + $^2$H $\lrt$ n + $^3$He and $^2$H + $^2$H $\lrt$ p + $^3$H were updated taking into account the data of \cite{Leonard:2006zm}.

After the first version of \texttt{PArthENoPE}, a revised version, \texttt{PArthENoPE 1.1}, was released at the beginning of 2017 \cite{parthenope_110}, where the default values of the physical parameters were updated to the  most recent experimental results (both in the configuration card and in the initialization routines) and the range for the reliable values of the neutron lifetime was changed to a 5 sigma range around the present experimental value quoted in \cite{PDG}. Besides this, an updated rate and uncertainty was implemented for the reaction $^2$H + p $\lrt$ $ \gamma$ + $^3$He, based on the  astrophysical factor given in \cite{Adelberger:2010qa}.

In the standard scenario, the only free parameter entering the BBN dynamics is the value of the baryon to photon number density, $\eta$. The relation of $\eta_{10}\equiv \eta\cdot 10^{10}$ with the present value of the baryon energy density, $\Omega_b h^2$, involves the $^4$He mass fraction, $Y_p$ (see e.g. \cite{Serpico:2004gx})
\be
\Omega_b h^2 = \frac{1-0.007125\, Y_p}{273.279} \left( \frac{T_{\gamma}^0}{2.7255 K} \right)^3 \eta_{10}
\ee
For this reason, \pthnew uses $\eta_{10}$ as input parameter and prints in the output file the corresponding value of $\Omega_b h^2$ using the value of $Y_p$ obtained in the run.

As \texttt{PArthENoPE}, also the new freely distributed version of \pthnew allows to treat non-standard physics through the choice of three parameters, which are general enough and widely used in the specialized literature:
\begin{enumerate}
\item
Extra degrees of freedom, $\dneff=N_{\rm eff} - 3.045$ \cite{deSalas:2016ztq}, (customarily referred to as the ``number of extra effective neutrino species"), related to the radiation energy density $\rho_X$ in non-electromagnetically interacting particles at the BBN epoch in addition to the three active neutrinos
\be
\rho_X=\frac{7}{8}\frac{\pi^2}{30}\dneff T_X^4\,,
\ee
where we assume $T_X$ to be equal to the neutrino temperature, $T_\nu$. This means that from the entropy conservation, $T_X=T$ at temperatures higher than the effective neutrino decoupling temperature, chosen as $T_d=2.3$ MeV, or else
\be
T_X=T\left[\frac{\hrho_{e\gamma B}(T)+
\hp_{e\gamma B}(T)}{\hrho_{e\gamma B}(T_d) +
\hp_{e\gamma B}(T_d)}\right]^{1/3}\,,\:\:T<T_d\:.
\ee
The user may input a value of $\dneff$ in the range $-3.0\leq\dneff\leq 3.0$.
\item
Chemical potential of the active neutrinos, $\xi_i$. In principle, the electron neutrino degeneracy parameter $\xi_e\equiv\mu_{\nu_e}/T_{\nu_e}$ as well as the degeneracy parameters of the other neutrino flavors are not determined within the cosmological model, and should be constrained observationally. They would be nonzero if a primordial lepton asymmetry exists in the form of neutrinos. If neutrinos were always kept in perfect kinetic and chemical equilibrium, the main impact of neutrino asymmetries on BBN would be a shift of the beta equilibrium between protons and neutrons, due to $\xi_e$, and a modification of the radiation density, due to all $\xi_i$
\be
\Delta N_{\rm eff}(\xi) = \sum_i\bigg[ \frac{30}{7} \bigg(
\frac{\xi_i}{\pi} \bigg)^2 + \frac{15}{7} \bigg( \frac{\xi_i}{\pi}
\bigg)^4 \bigg]\,. \label{e:dnxi}
\ee
However, it was shown in a series of papers (see e.g.\ \cite{Mangano:2011ip,Castorina:2012md}) that the relations between the lepton asymmetry (and $N_{\rm eff}$) and the neutrino degeneracy parameters $\xi_i$ in equilibrium do not necessarily hold due to the combined effect of flavour oscillations and collisions around the epoch of neutrino decoupling. In any case, for the
values of the neutrino mixing parameters preferred by global fits of oscillation data \cite{Capozzi:2017ipn,Esteban:2016qun,deSalas:2017kay}, and in particular that of $\sin^2 \theta_{13}$, the impact of a potential lepton asymmetry can be approximated by choosing a common value of the degeneracy parameters, 
$\xi \simeq \xi_{\mu} \simeq \xi_{\tau} \simeq \xi_e$. 
In the present version of \texttt{PArthENoPE}, we give the possibility to the users to explore alternative scenarios with any value of $\xi_e$ and $\xi_x\equiv\xi_{\mu,\tau}$, and values in the range $[-1,1]$ are allowed.\footnote{In the previous version of the code, the user could only select a discrete grid of values of $\xi\equiv\xi_e=\xi_{\mu,\tau}$.}
\item
Energy density of the cosmological constant, $\rho_\Lambda$, which enters the equations only via its contribution to the Hubble parameter
\be
H = \sqrt{\frac{8\,\pi\, G_N}{3} \left(\rho+\rhol \right)}\:.
\ee
with allowed range $0\leq(\rhol$/MeV$^4)\leq 1$.
\end{enumerate}

Of course, other non-standard physics scenarios can be also considered by intervening on the source code (such as non-thermal neutrino distributions, quintessence fluids, photon injection, etc.)

%%%%%%%%%%%%%%%%%%%%%%%%%%%%%%%%%%%%
\section{The new release of \pth}
\label{new release} 
In this new release, the \pth code consists of the Fortran files \texttt{main2.0.f} and \texttt{parthenope2.0.f}, already present in the old distribution, some auxiliary Fortran files with the new routines used in the code (see below), and one Python file, \texttt{gui2.0.py}.

The numerical method used for solving the BBN Ordinary Differential Equations (ODE) in the old versions of the code, until \texttt{PArthENoPE 1.1}, was Backward Differentiation Formulas implemented in a NAG routine. While \texttt{PArthENoPE 1.1} remains still available upon request of the interested users, an effort was made to replace the NAG libraries, specialized for the resolution of complex mathematical problems but, unfortunately, not free, with other libraries publicly available.

Beside this, we introduced a more user friendly GUI, which enhances the old \texttt{main.f} interface of the previous version of \texttt{PArthENoPE}, for choosing interactively the initial parameters and running configuration. This means that in the new distribution of \pth the user can choose between the old command line mode based on the two Fortran files and configuration file or the graphical interface contained in the Python file \texttt{gui2.0.py}.

In the following, we describe the two main improvements in the present version of the code.

%%%%%%%%%%%%%%%%%%%%%%%%%%%%%%%%%%%%
\subsection{\pthnew}
\label{ss:code}
%%%%%%%%%%%%%%%%%%%%%%%%%%%%%%%%%%%%

We address the reader to Sec.\ 4 of \cite{parthenope} for the description of the routines contained in the file \texttt{parthenope2.0.f}. In this regard, we can make the following comments:
\begin{itemize}
\item
We kept the old name \texttt{fcn} of the routine that calculates the right-hand side of the differential BBN equations, which has been changed to the format requested by the new solver.
\item
The weak rates are now calculated in a separate routine, \texttt{wrate}, called by the routine \texttt{rate} where all reaction rates are evaluated. This makes easier for the user to adopt, if needed, a user-defined routine to compute weak rates in non-standard physics scenarios.
\item
We increased the dimension of the common OUTVAR, which now contains free locations (from 4 to 20) for user-defined output variables. In order to use this option, the user needs to add in the file \texttt{parthenope2.0.f} the necessary instructions to store the desired quantities in the array OUTVAR (wherever they are calculated), their names (as string variables) in the corresponding array OUTTXT (in the subroutine INIT) and update the integer NOUT (in the subroutine INIT), whose value indicates the number of output variables to be printed.
\end{itemize}
Several mathematical operations were computed by the NAG routines in the old version of \texttt{PArthENoPE}, among them the most important one was the resolution of the coupled BBN ODE by the Backward Differentiation Method. We list in the following the changes made in the algorithms used inside \texttt{PArthENoPE}:
\begin{itemize}
\item
The implicit equation~\eqn{e:L} is now solved by using Brent's method to find the root of a function, previously bracketed with inward search in an initial interval: the code used is based on the \texttt{zbrak} and \texttt{zbrent} routines in \cite{numericalrecipes}.
\item
The one-dimensional integration needed for calculating the function $\lh \left(m_e/T,\pe\right)$ in eq.~\eqn{e:lfunc} is performed by an extended trapezoidal rule: the code used is based on the \texttt{qtrap} and \texttt{trapzd} routines in \cite{numericalrecipes}.
\item
Modified Bessel functions of the second kind are necessary in the code for evaluating several electron quantities, like for example $\rho_e$ and $p_e$: to compute these functions the code now uses \texttt{bessi0}, \texttt{bessk0}, \texttt{bessi1}, \texttt{bessk1}, \texttt{bessk} from \cite{numericalrecipes}.
\item
In \pthnew the BBN ODE are solved using the routine \texttt{DLSODE}, included in the package \texttt{ODEPACK} \cite{odepack}. This routine was selected among the ones available in \texttt{ODEPACK} because it solves nonstiff or stiff systems.
\end{itemize}

While the routines based on \cite{numericalrecipes} are included in the file \texttt{addon2.0.f}, the auxiliary routines called for the ODE system resolution are contained in the files \texttt{odepack1-parthenope2.0.f} and \texttt{odepack2-parthenope2.0.f}. Note that some unused routines contained in the original files \texttt{odepack\_sub1.f} and \texttt{odepack\_sub2.f} of the \texttt{ODEPACK} distribution were deleted in order to avoid some error messages that would have caused troubles for the users. We have also fixed a problem of an underflow exception and introduced few other minor modifications.

\subsection{The \pth Graphical User Interface}

In the old version of the code, the user had the possibility of choosing between two modes, an interactive (but lengthy) one, and a card mode, faster but less user friendly due to the requirement of using a precise syntax based on given keywords. The interface between the user and the BBN code contained in the file \texttt{parthenope.f} were included in the file \texttt{main.f}. The GUI does not substitute this interface, which was conserved for backward compatibility, but improves it, since it allows to produce in a very simple way the configuration file (corresponding to the input card in the `card mode' of the old version). In particular, in addition to the possibility of running the code with single values of the physical parameters, the GUI allows the user to send multiple jobs of \pth with the physical parameters varying on a grid.

The GUI of \pth is implemented in Python. To run the GUI the user needs a Python version 2.7 as well as application dependencies (libraries and modules): Tkinter, Pmw, ttk, Numpy and PIL\footnote{For details on the installation and configuration of the necessary software, see the GUI user manual on the \pth web page \cite{parthenope}.}. 

The GUI is structured in overlapping pages and  consists of four main sections. Each of the first three sections is dedicated to a specific set of parameters and/or features that can be chosen by the user. More precisely, the user can choose the nuclear network, the physical input parameters and customize the output. The fourth section allows the running of the code with the chosen parameters.

The start page of the GUI provides the user with useful links for general information and four buttons for accessing each of the above mentioned sections, where specific widgets, which show the allowed range of values, help the user in the choice of the input parameters and/or configuration. As an alternative it is possible to directly open the run page for launching \pth with default values of all parameters, that is the default nuclear network and physical input values (single values). 

The section ``Network parameters'' allows the user to choose among the small, intermediate or complete network corresponding, respectively, to 9 nuclides and 40 reactions associated,  18 nuclides and 73 reaction associated, 26 nuclides and 100 reactions associated. The user can change the rate of any number of such reactions by accessing the ``Nuclear reactions" page that opens clicking on the corresponding button. There, it is possible to select the reactions and the type of change to be applied to their rates, using a \textsl{drop-down list} and a \textsl{counter} widget (i.e.\ an entry field with arrow buttons to increase and decrease the current value or type a new one in the allowed range). In particular, each rate can be modified by $\pm \sigma$, or a multiplicative factor choosing `High', `Low', or `Factor', respectively. Low and high values correspond to the variation by the rate uncertainty, whereas the multiplicative factor can be chosen in the range [0, 10]. A summary window can be opened from the ``Network parameters'' page, to display the choices made, and to possibly implement additional reactions rate changes going back to the ``Nuclear Reactions" page.

The section ``Physical parameters'' allows the user to choose a single value or a grid of values for the baryon to photon number density, effective number of neutrinos, neutron lifetime, neutrino chemical potentials, and energy density of the vacuum at the BBN epoch.  The user can set the grid by entering the number of grid points, $N$, in the entry widget below the corresponding buttons. Optionally, it is possible to restrict the interval to a smaller one between the minimum and maximum  values indicated. In case the user prefers to choose single values for some or all of the physical parameters, the entry field of the ``$N$'' button has to be left equal to zero.

The section ``Output options'' is devoted to the choice of the name of the output files, the first one with the final results and the second one with the evolution of the abundances of the selected nuclides. Such nuclides can be chosen by a \textsl{drop-down list} and visualized on a window opened by the button ``Nuclides''. 

Finally, the section ``Run PArthENoPE'' allows to launch the code with the default values or with the values of the parameters implemented by the user. When the corresponding button is clicked, the GUI produces one or more configuration files, needed for \pth running, depending on the choice of single values or grid of values of the parameters. Each configuration file is connected to the corresponding pair of output files through the date and time of the run and an identifying number. All configuration and output files produced in a given run of the GUI are moved to a subdirectory of the working directory, whose name contains the values of date and time that identify the run. Moreover, in this directory the user can find the file $grid_{-}date_{-}time.out$ where all the final values of the abundances of the selected nuclides are reported, in order to make easier to produce plots and perform analyses.

The user is supported both in the usage of widgets (buttons, drop-down lists, text boxes, counters) and allowed choices of parameters by a quick help guide provided on each page. Further details on the GUI can be found on the GUI user manual distributed in the \pthnew package, obtainable on the web page of \texttt{PArthENoPE}.

\begin{figure}[p]
\includegraphics[width=\textwidth]{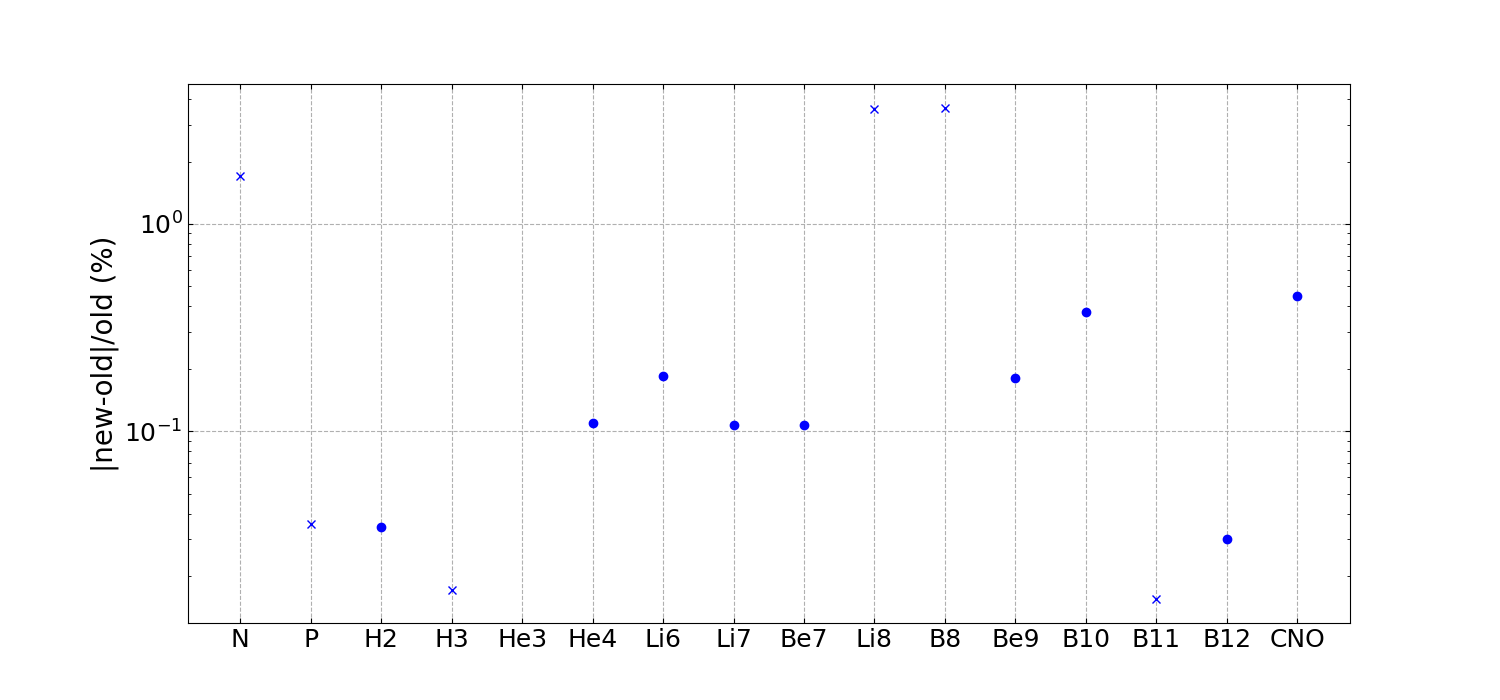}
\caption{Absolute value of the relative difference between versions (in percentage) of the final primordial abundances for all nuclides computed in \texttt{PArthENoPE}. CNO includes all carbon, nitrogen and oxygen nuclides. Circles are positive relative differences and crosses correspond to negative differences. The difference for $^3$He/H is smaller than 0.001\%.}
\label{fig:rel-abs-diff-all}
\end{figure}

\begin{figure}[p]
\includegraphics[width=.8\textwidth]{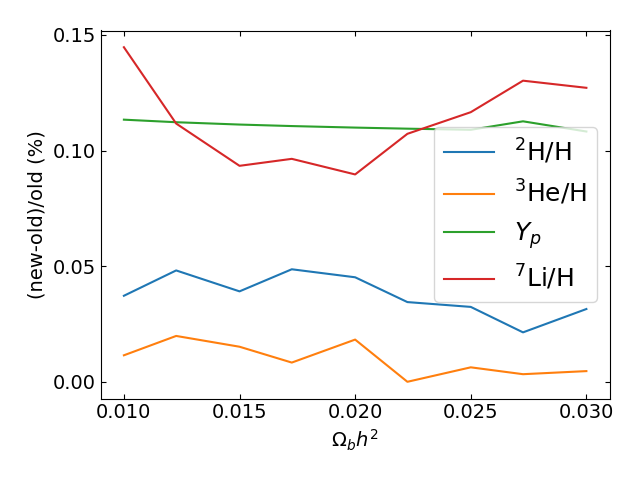}
\caption{Relative difference between \pthnew and \pth (in percentage) of the produced $^{2}{\rm H/H}$, $^3$He/H, $Y_p$ and $^7$Li/H for different values of the
baryon density $\Omega_b h^2$.}
\label{fig:H2He4vsOmbh2-diff}
\end{figure}

\begin{figure}[p]
\includegraphics[width=.8\textwidth]{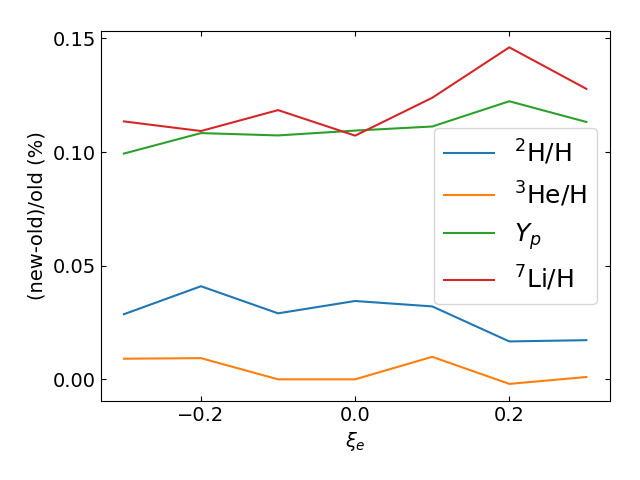}
\caption{Same as Fig. \ref{fig:H2He4vsOmbh2-diff} but varying the neutrino degeneracy parameter with $\xi_e=\xi_\mu=\xi_\tau$.}
\label{fig:H2He4vsXie-diff}
\end{figure}

\begin{figure}[p]
\includegraphics[width=.8\textwidth]{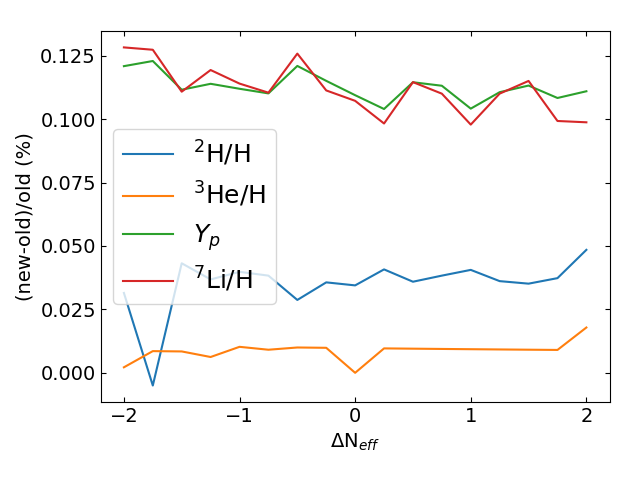}
\caption{Same as Fig. \ref{fig:H2He4vsOmbh2-diff} but varying the effective number of neutrinos ($\Delta N_{\rm eff}$).}
\label{fig:H2He4vsNnu-diff}
\end{figure}

%%%%%%%%%%%%%%%%%%%%%%%%%%%%%%%%%%%%%%%%
\section{Comparison with \pth}
\label{s:compare}

We have checked that the performance of \pthnew is at least as good as that of \pth and we can recover a nice evolution of the different nuclides. For this purpose we compared the outputs of both versions using a benchmark model in which we choose the standard values  $\Delta N_{\rm eff}=0$, $\eta_{10} = 6.09389$ (corresponding to $\Omega_b h^2 = 0.02226$ \cite{Ade:2015xua}) and $\xi_i = 0$.

In Fig.~\ref{fig:rel-abs-diff-all} we show the relative difference of the final abundances of all nuclides in percentage. The total abundance of carbon, nitrogen and oxygen nuclides is shown in a single column labelled CNO, i.e.\ the total metallicity produced during BBN, which is very small (of order $10^{-16}$) and largely dominated by $^{12}$C, see e.g.\ \cite{Iocco:2007km}. As can be seen in the figure, the difference is below 1\% for most of the nuclides; in particular, for deuterium and helium it is below $\sim 0.1$\%.

In order to show the robustness of the new code beyond our benchmark model, we checked its behaviour in the parameter space when leaving free one of the three fixed parameters stated above. We show in Figures~\ref{fig:H2He4vsOmbh2-diff}, \ref{fig:H2He4vsXie-diff} and \ref{fig:H2He4vsNnu-diff} the relative difference between the output abundances of the four main nuclides  from \pth and \pthnew  for different values of $\eta_{10}$ (for user convenience we report in the plot the values of $\Omega_b h^2$), $\xi_e=\xi_\mu=\xi_\tau$ and $\Delta N_{\rm eff}$, respectively. As can be seen in the plots, no relevant dependence on the value of the parameters is found, with the exception of $^7$Li as a function of the baryon density, but for some value of $\Omega_b h^2$ excluded by Planck results.

Note that the number of steps performed by the ODE system solver in \pth on each simulation is about 5 times that of \texttt{PArthENoPE 2.0}. However, as proved by the results shown in the above figures, this does not change the precision of the code and the difference in the number of steps is related with the internal construction of the different solvers.

\section{BBN analysis using \pthnew}
\label{s:analysis}

\begin{figure}[t]
\begin{center}
\includegraphics[width=0.6\textwidth]{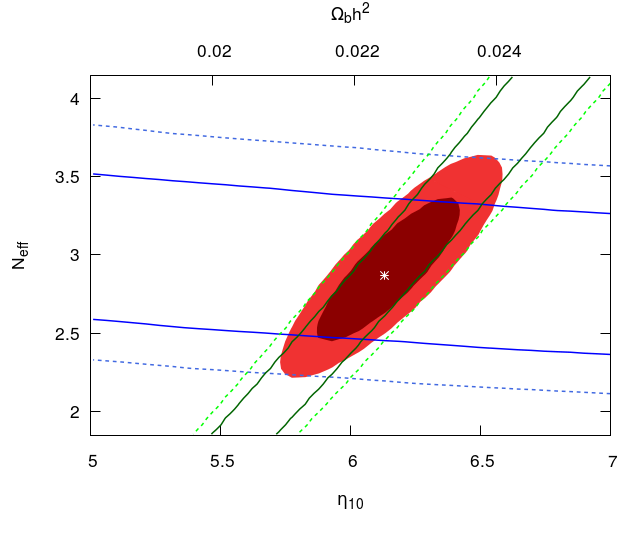}
\caption{Allowed regions for the cosmological baryon density and the effective number of neutrinos from our BBN analysis, shown as contours at the $1\,\sigma$ and $2\,\sigma$ levels. For comparison, we also include the corresponding intervals for the experimental measurements of $^4$He (blue lines) and deuterium (green lines) alone.}
\label{fig:analysis-Yp-D}
\end{center}
\end{figure}

In this section we illustrate the potentiality of the new version of \texttt{PArthENo} \texttt{PE} performing a BBN analysis, both in the standard case with a single free parameter ($\eta_{10}$) and in an extended scenario where the radiation content of the universe can vary via the parameter $N_{\rm eff}$. The theoretical output of the primordial abundances of the light nuclei from our code is compared with recent experimental data. Here we consider two recent measurements on the abundances of the main nuclides: deuterium \cite{Cooke:2017cwo} and $^{4}\mathrm{He}$ \cite{Aver:2015iza} (we treat all uncertainties to be $1\sigma$ throughout this section),
\begin{equation}
^{2}\mathrm{H/H} = \left( 2.527\pm 0.030 \right) \cdot 10^{-5}
\end{equation}
and
\begin{equation}
Y_p = 0.2449\pm 0.0040\pp
\end{equation}
We used the abundances of both elements to constrain $\eta_{10}$ and $N_{\rm eff}$ defining a $\chi^2$-function as follows
\begin{equation}
\chi^2 = \sum_i \frac{\left( X_i^{\rm th}\left( N_{\rm eff}, \eta_{10} \right) - X_i^{\rm exp} \right)^2}{\sigma^2_{i,{\rm th}}+\sigma^2_{i,{\rm exp}}},
\end{equation}
where $i =\{\, ^{2}\mathrm{H/H}\, ,\, Y_p\}$, $\sigma_{i,{\rm exp}}$ are the experimental errors as above and $\sigma_{i,{\rm th}}$ is the uncertainty due to propagation of nuclear process rates: $\sigma_{^{2}\mathrm{H/H},{\rm th}}=0.05\cdot 10^{-5}$ and $\sigma_{Y_p,{\rm th}}=0.0003$ \cite{Coc:2015bhi}.

When the effective number of neutrinos is fixed to its standard value, $N_{\rm eff}=3.045$, from our analysis we find that the baryon to photon number density is $\eta_{10}=6.23^{+0.12}_{-0.14}$. This corresponds to a present baryon density allowed by BBN of $\Omega_b h^2=0.0227\pm 0.0005$, in excellent agreement with the range $\Omega_b h^2 = 0.02218\pm 0.00015$ measured by Planck \cite{Aghanim:2016yuo}. Instead, the $^7$Li abundance from \pthnew for the best-fit value of $\eta_{10}$ is $^7{\rm Li/H}=4.69\cdot 10^{-10}$. This value clearly disagrees with the measurement $^{7}\mathrm{Li/H} = (1.58\pm0.31) \cdot 10^{-10}$  \cite{Sbordone:2010zi}, the already mentioned primordial lithium discrepancy.

The results of our BBN analysis when $N_{\rm eff}$ is left as a free parameter can be seen in Fig.~\ref{fig:analysis-Yp-D}. The allowed regions for $N_{\rm eff}$ and  $\eta_{10}$ are shown at the 1 and $2\sigma$ levels, together with the bounds when only the experimental measurement of deuterium or $^4$He is considered. The corresponding allowed ranges are 
\begin{equation}
N_{\rm eff}=2.87^{+0.24}_{-0.21}\qquad {\rm and} \qquad \eta_{10}=6.13\pm 0.13~ (\Omega_b h^2=0.0224\pm 0.0005)
\label{BBNresults}
\end{equation}
Both agree with the radiation content expected in the standard case ($N_{\rm eff}=3.045$) and the baryon density favoured by the analysis of data on CMB anisotropies by Planck \cite{Aghanim:2016yuo}, respectively. Finally, we have included in our analysis the Planck measurement for $\Omega_b h^2$ as a gaussian prior. In this case, we obtain the allowed ranges $N_{\rm eff} = 2.81^{+0.22}_{-0.21}$ and $\eta_{10} = 6.082^{+0.022}_{-0.06}$ ($\Omega_b h^2 = 0.02224^{+0.00008}_{-0.00020}$). 

\section{Conclusions}
\label{s:concl}

CMB anisotropies, large-scale structures, gravitational lensing and BBN are an intertwined system of observables which test the cosmological evolution at different epochs. The quantity and quality of data coming from astrophysical measurements in the last decade, and those which will be available in the future, call for accurate tools providing theoretical predictions on cosmological observables to check the overall
consistency of the picture of the evolution of the universe, as well as for investigating and constraining new physics beyond the present framework of fundamental interactions. In this perspective, in 2008 we developed the BBN numerical code \pth with two main goals: to create a tool with an increased level of accuracy of theoretical predictions on nuclide abundances (improved estimate of the neutron to proton weak conversion rates, detailed neutrino decoupling, reanalysis of nuclear network rates) which updated the works of \cite{Wagoner:1966pv,Wagoner2,Wagoner:1972jh,Kawano:1988vh,Smith:1992yy}, and to provide a public tool to researchers interested in the field which fulfills the requirements of precision and versatility. After almost ten years we have developed a new version of the code, \texttt{PArthENoPE 2.0}, whose new aspects and characteristics have been described in this paper. Apart from minor points, which will be detailed at the \pth web page, {\texttt{http://parthenope.na.infn.it}},  we have here outlined the main new features of \texttt{PArthENoPE 2.0}:

\begin{itemize} 
\item We replaced the NAG libraries used for various purposes in the code, but which are not free software, with other libraries publicly available. 
\item \pthnew is now interfaced with a Python file, which enhances the old \texttt{main.f} interface of the previous version and  provides a more user friendly GUI for choosing interactively the initial parameters and running configuration. 
\end{itemize}
\bigskip
We hope this new version of the code, which shares with \pth the same theoretical and numerical accuracy, but is more user friendly, will be useful to researchers interested in BBN-related studies.

\section*{Acknowledgments}
We are grateful to many users of \texttt{PArthENoPE}, who during these years, gave important feedbacks and suggestions to improve the code. 
This work was supported by INFN, under Iniziativa Specifica TASP, and by {\it Fondi della regione Campania}, ``L.R. num. 5/2002 -- annualit\`a 2007". 
P.F.d.S.\ and S.P.\  were supported by the Spanish grants FPA2017-85216-P and SEV-2014-0398 (MINECO), PROMETEOII/2014/084 (Generalitat Valenciana) and FPU13/03729 (MECD).

\appendix

\section{The \pth set of equations}
\label{ap:erratum}

Few typos and minor mistakes\footnote{We thank Ahmad Borzou for pointing out these typos.} are contained in the Appendix A of the \pth reference paper \cite{parthenope}. In this Appendix we summarize the \pth set of equations with the purpose of correcting these typos. We remind the reader that we use the dimensionless 
variables $z\equiv m_{e}/T$, $x=m_{e}\,a$, $\bar{z}=x/z=a T=T/T_\nu$. Moreover, as defined in \cite{parthenope},
\bea
&{\mathcal N}(z)=\frac{1}{\bar{z}^4}\left(x
\frac{d}{dx}\bar\rho_\nu\right)\Bigg|_{x=x(z)}\,\,\,,  &
\bar{\rho}_{\nu}=a^4\,\rho_{\nu}=\left(\frac{x}{m_e}\right)^4\,\rho_{\nu}
\,\,\,,
\label{e:Nz}\\
&\rho=\rho_{e \gamma B}+\rho_\nu \,\,\,,  &\p=\p_{e \gamma B}+\p_\nu \vv \\
&\hrho=T^{-4}\,\rho=\left(\frac{z}{m_e}\right)^4\,\rho \vv
&\hp=T^{-4}\,\p=\left(\frac{z}{m_e}\right)^4\,\p \pp
\label{e:rhophat}
\eea
Finally, in what follows we will also consider the energy density of electrons/positrons plus photons, $\rho_{e \gamma}$. The equation for the entropy conservation
\be
\dot{\rho} = -\, 3 \, H~ (\rho + \p) \vv 
\ee
once used the definitions \eqn{e:Nz}-\eqn{e:rhophat} and expressing the total time derivative of $\hrho_{e \gamma B}$ via the partial derivatives with respect to $z$, $\pe$ and $X_i$, becomes
\be
\left( \frac{\partial \hrho_{e \gamma B}}{\partial z} -
\frac{4}{z}\,\hrho_{e \gamma B} + \frac{\partial \hrho_{e \gamma
B}}{\partial \pe} \, \frac{d\pe}{dz}  +\sum_i \frac{\partial
\hrho_{e \gamma B}}{\partial {X_i}} \, \frac{dX_i}{dz} \right)
\dot{z} = -\,3\,H\,(\hrho_{e \gamma B} + \hp_{e \gamma
B})-H\,\mathcal{N}(z)\pp \label{e:dotz1}
\ee
Proceeding in the same way, the equation for conservation of the total baryon number
\be
\dot{n}_B = -\, 3\, H\, n_B \vv
\ee
becomes
\be
\left(\frac{\partial \hnb} {\partial z}+ \frac{\partial \hnb}
{\partial \pe} \frac{d\pe}{dz}+\sum_i \frac{\partial \hnb}
{\partial {X_i}} \frac{dX_i}{dz} \right) \dot{z} = - \, 3\,H \, \hnb \pp
\label{e:dotz2}
\ee
where we define $\hnb \equiv n_B/m_e^3= \lh(z,\pe)/(z^3 \sum_i Z_i X_i)$. Obtaining $\dot{z}$ from~\eqn{e:dotz2} and substituting into~\eqn{e:dotz1} we get
\be
-\,3\,H\,\hnb\frac{ \frac{\partial \hrho_{e \gamma B}}{\partial
z}-\frac{4}{z}\hrho_{e \gamma B}+ \frac{\partial \hrho_{e \gamma
B}}{\partial \pe}\,\frac{d\pe}{dz} +\sum_i \frac{\partial \hrho_{e
\gamma B}}{\partial {X_i}} \,\frac{dX_i}{dz}}{ \frac{\partial
\hnb}{\partial z}+\frac{\partial \hnb}{\partial \pe}
\frac{d\pe}{dz}+\sum_i \frac{\partial \hnb}{\partial {X_i}}
\frac{dX_i}{dz}}= -\,3\,H\,(\hrho_{e \gamma B} + \hp_{e \gamma
B})-H\,\mathcal{N}(z) \pp
\label{e:eq3}
\ee
By solving this equation with respect to $d\pe/dz$ (expressing $\hnb$ and its derivatives as function of
$\lh(z,\pe)$) one gets the expression for $d\pe/dz$ used in \pth
\be
\frac{d\pe}{dz} = \frac1z \frac{\lh\, \kappa_1 +\left(\hrho_{e
\gamma} + \hp_{e \gamma B}+
\frac{\mathcal{N}(z)}{3}\right)\,\kappa_2}{\lh\, \frac{\partial
\hrho_e}{ \partial \pe} -\frac{\partial \lh}{
\partial \pe} \left(\hrho_{e \gamma} + \hp_{e \gamma B}+
\frac{\mathcal{N}(z)}{3}\right)}\vv
\label{e:dphidz}
\ee
where
\bea
\kappa_1 &=& 4\, \left(\hrho_{e}+ \hrho_{\gamma}\right) + \frac32~
\hp_B - z\, \frac{\partial \hrho_e}{\partial z} -z\,
\frac{\partial \hat{\rho}_\gamma}{\partial z}
-\frac{z^2\, \lh}{\sum_j Z_j\, X_j}\sum_i \left(
\hdmi + \frac{3}{2\, z} \right)\tgi \vv \nonumber \\
&& \\
\kappa_2 &=& z\,\frac{\partial \lh}{\partial z}-3\,\lh -
z\,\lh\,\frac{ \ds \sum_i~ Z_i\, \tgi}{\ds \sum_j~ Z_j\, X_j}\pp
\eea
In the previous equations, $\hdmi$ stands for the i-th nuclide mass excess normalized to $m_e$, whereas $\tgi\equiv d\Gamma_i/dz$.

We now need $\dot{z}$ in order to calculate $\tgi=\Gamma_i/\dot{z}$. At this aim, one can substitute Eq.~\eqn{e:dphidz} in \eqn{e:dotz2}, getting after some rearrangements
\be
\dot{z}= -3\, z\, H\, \frac{\lh\, \frac{\partial
\hrho_e}{ \partial \pe} -\frac{\partial \lh}{
\partial \pe} \left(\hrho_{e \gamma} + \hp_{e \gamma B}+
\frac{\mathcal{N}(z)}{3}\right)}{\kappa_1 \, \frac{\partial \lh}{\partial
\pe} +\kappa_2 \, \frac{\partial \hrho_e}{\partial
\pe}}\pp
\ee
The equations for the abundances~\eqn{e:dXdt} then assume the form used in \texttt{PArthENoPE},
\be
\frac{dX_i}{d z}= -\frac{\hgi}{3\,
z\,\hH}\, \frac{\kappa_1 \, \frac{\partial \lh}{\partial \pe}
+\kappa_2 \, \frac{\partial \hrho_e}{\partial
\pe}}{\lh\,\frac{\partial\hrho_e}{\partial\pe}-\frac{\partial
\lh}{\partial \pe}\left(\hrho_{e \gamma} + \hp_{e \gamma B} +
\frac{\mathcal{N}(z)}{3}\right) }\pp
\ee

\end{document}